\documentclass[conference]{IEEEtran}
\IEEEoverridecommandlockouts
\usepackage{cite}
\usepackage{amsmath,amssymb,amsfonts}
\usepackage{algorithmic}
\usepackage{graphicx}
\usepackage{comment}
\usepackage{textcomp}
\usepackage{amsthm}
\usepackage{xcolor}
\usepackage{bbding}
\usepackage{makecell}
\usepackage{algorithm}  
\usepackage{color}
\usepackage{multirow}
\usepackage{bm}
\usepackage{array}
\usepackage{booktabs}
\usepackage{tabularx,booktabs}
\usepackage{multicol}
\usepackage{cleveref}
\usepackage{balance}
\usepackage{lastpage}
\usepackage{tabulary}
\usepackage{subfigure}
\usepackage{etoolbox}
\usepackage{bbm}
\usepackage{enumitem}
\usepackage{mathrsfs}
\usepackage{multicol}
\usepackage{amsfonts,amssymb}
\usepackage{setspace}
\usepackage{ulem}
\usepackage{stackengine}
\usepackage{fancyhdr} 
\usepackage{threeparttable}

\usepackage{amsthm}
\theoremstyle{definition}

\def\BibTeX{{\rm B\kern-.05em{\sc i\kern-.025em b}\kern-.08em
    T\kern-.1667em\lower.7ex\hbox{E}\kern-.125emX}}

\renewenvironment{IEEEbiography}[1]
  {\IEEEbiographynophoto{#1}}
  {\endIEEEbiographynophoto}

\begin{document}

\title{Heuristic Algorithms for RIS-assisted \\ Wireless Networks: Exploring Heuristic-aided Machine Learning  
\thanks{
H. Zhou and M. Erol-Kantarci are with the School of Electrical Engineering and Computer Science, University of Ottawa, Ottawa, ON K1N 6N5, Canada. (emails:\{hzhou098, melike.erolkantarci\}@uottawa.ca).

Yuanwei Liu is with the School of Electronic Engineering and Computer Science, Queen Mary University of London, London E1 4NS, U.K. (email:yuanwei.liu@qmul.ac.uk).

H. Vincent Poor is with the Department of Electrical and Computer Engineering,
Princeton University, Princeton, NJ 08544 USA (email: poor@princeton.edu).} 
}

\author{\IEEEauthorblockN{Hao Zhou, Melike Erol-Kantarci, \IEEEmembership{Senior Member, IEEE}, Yuanwei Liu, \IEEEmembership{Senior Member, IEEE},\\ and H. Vincent Poor, \IEEEmembership{Life Fellow, IEEE}} }

\maketitle

\thispagestyle{fancy}            
\chead{This paper has been accepted by IEEE Wireless Communication Magazine. } 

\renewcommand{\headrulewidth}{1pt}      
\pagestyle{plain}

\begin{abstract}
Reconfigurable intelligent surfaces (RISs) are a promising technology to enable smart radio environments. However, integrating RISs into wireless networks also leads to substantial complexity for network management. This work investigates heuristic algorithms and applications to optimize RIS-aided wireless networks, including greedy algorithms, meta-heuristic algorithms, and matching theory. Moreover, we combine heuristic algorithms with machine learning (ML), and propose three heuristic-aided ML algorithms, namely heuristic deep reinforcement learning (DRL), heuristic-aided supervised learning, and heuristic hierarchical learning. Finally, a case study shows that heuristic DRL can achieve higher data rates and faster convergence than conventional deep Q-networks (DQN). This work provides a new perspective for optimizing RIS-aided wireless networks by taking advantage of heuristic algorithms and ML.
\end{abstract}


\section{Introduction}
Reconfigurable intelligent surfaces (RISs) have emerged as an attractive technology for envisioned 6G networks, enabling smart radio environments to improve energy efficiency, network coverage, channel capacity, and so on\cite{yliu}. 
Many existing studies have demonstrated the capability of RISs, but incorporating RISs into existing wireless networks significantly increases the network management complexity. In particular, each small RIS element requires sophisticated phase-shift control, resulting in a very large solution space.         
In addition, RISs can be combined with other technologies due to low hardware costs and energy consumption, such as unmanned aerial vehicle (UAV) and non-orthogonal multiple access (NOMA), leading to joint optimization problems with coupled control variables. 
Therefore, efficient optimization techniques are crucial to realizing the full potential of RISs.

Existing RIS optimization techniques can be categorized into three main approaches: convex optimization, machine learning (ML), and heuristic algorithms \cite{zhou2023survey}. Convex optimization algorithms have been widely applied for optimizing RIS-aided wireless networks, e.g., using block coordinate descent for joint active and passive beamforming, and applying fractional programming to decouple the received signal strength with interference and noise \cite{cunh}. 
However, RIS-related optimization problems are usually non-convex and highly non-linear, and then the application of convex optimization requires case-by-case analyses, resulting in considerable difficulties for algorithm designs. 

By contrast, ML algorithms have fewer requirements for problem formulations. For instance, when using a Markov decision process (MDP) for RIS passive beamforming, channel state information (CSI), RIS phase shifts, and objective functions are defined as states, actions, and rewards, respectively.
Then, reinforcement learning algorithms are deployed to maximize the reward by trying different action combinations. 
However, ML algorithms are usually computationally demanding, which means many iterations and tedious model training. The low training efficiency may hamper the system efficiency and result in slow convergence \cite{zhou2022knowledge}.   

Different from convex optimization and ML algorithms, heuristic algorithms obtain low-complexity solutions efficiently, trading optimality and accuracy for speed. There are multiple motivations for investigating heuristic algorithms in RIS-aided wireless networks. 
Firstly, heuristic policies can be easily applied to various scenarios with few requirements on problem formulations. Such high generalization capability indicates that heuristic algorithms can adapt well to different RIS-aided network applications, e.g., UAV-RIS and NOMA-RIS systems \cite{liu2022reconfigurable}.   
Secondly, heuristic algorithms are usually computationally efficient, and therefore they can respond rapidly to real-time network dynamics. 
Moreover, nondeterministic polynomial (NP)-hard problems often arise in RIS-aided networks, such as discrete phase-shift design and resource allocation problems, and heuristic algorithms are particularly useful for solving these problems.
Despite the advantages, heuristic rules produce only locally optimal solutions, e.g., particle swarm optimization (PSO) usually produces locally optimal results. Therefore, sub-optimality is considered to be the main issue with heuristic algorithms. However, these heuristic rules can be combined with other techniques, such as ML, offering a new opportunity for developing low-complexity and efficient algorithms.

Convex optimization and ML approaches have been investigated by a few studies for optimizing RIS-aided networks. For example, Liu \textit{et al.} overviewed joint beamforming and resource management problems \cite{yliu}, Pan \textit{et al.} summarized optimization techniques on signal processing for RIS-aided wireless systems \cite{cunh}, and Faisal \textit{et al.} focused on ML techniques for RIS-related applications \cite{kfai}. 
In these existing studies, heuristic algorithms are usually considered as baselines or supplements. 
However, due to the low-complexity feature, heuristic algorithms may provide a new perspective for optimizing RIS-aided networks. Moreover, heuristic algorithms can be combined with other techniques for joint optimization, enabling higher flexibility for RIS control. Our former work \cite{zhou2023survey} summarized model-based, heuristic, and ML approaches for optimizing RIS-aided wireless networks, but combining heuristic algorithms with ML was not investigated, which provides one of the main motivations of this work.  

The main contributions are summarized as follows: 
\begin{itemize}
    \item Unlike existing studies, this work comprehensively analyzes heuristic algorithms and their applications to RIS-aided wireless networks. We investigate existing heuristic algorithms, such as greedy algorithms, meta-heuristic algorithms, and matching theory, in terms of features, advantages, disadvantages, and difficulties. It provides a systematic overview of how to apply heuristic algorithms to RIS-aided wireless networks.   
    \item We combine heuristic algorithms with ML approaches, and propose three novel heuristic-aided ML techniques: 1) combining greedy algorithms with deep reinforcement learning (DRL) to accelerate DRL model training; 2) using heuristic algorithms to produce fine-grained datasets for supervised learning; and 3) deploying heuristic algorithms in hierarchical learning for fast decision-making. Such combinations offer new opportunities for optimizing RIS-aided wireless networks. Finally, a case study demonstrates that heuristic ML can achieve faster convergence and better network performance than conventional ML.  
\end{itemize} 

\begin{figure*}[!t]
\centering
\includegraphics[width=0.95\linewidth]{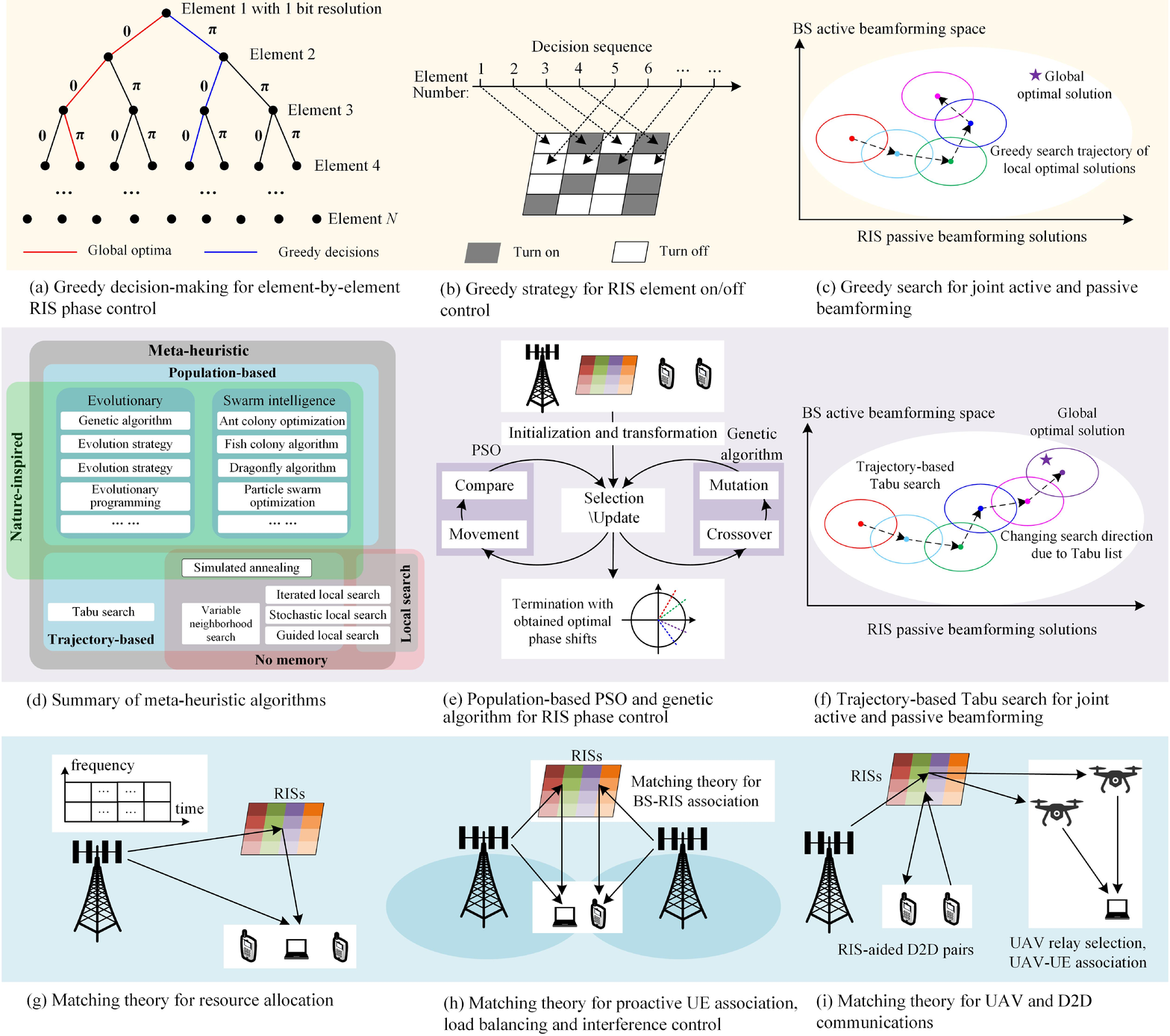}
\setlength{\abovecaptionskip}{-4pt} 
\caption{Heuristic algorithms for optimizing RIS-aided wireless networks.}
\label{fig-heu}
\vspace{-14pt}
\end{figure*}

\section{Heuristic Algorithms for Optimizing RIS-aided Wireless Networks}
\label{sec-heu}
This section provides a systematic overview of greedy algorithms, meta-heuristic algorithms, and matching theory, including their features, advantages, difficulties, and applications. Such a comprehensive overview is crucial for properly applying these heuristic algorithms to RIS-aided networks

\subsection{Greedy Algorithms}
Greedy algorithms have been widely used as heuristic optimization approaches in many problems, and Fig. \ref{fig-heu} shows the application of greedy algorithms to RIS-related optimization problems. 
Phase-shift optimization is the core of RIS control, and one of the main challenges is that the phases of all RIS elements are usually jointly optimized to achieve global optimality. Instead of complicated joint optimization, greedy algorithms are applied as low-complexity alternative approaches. 
As shown in Fig. \ref{fig-heu}(a), it decides the optimal phase shift of one element at each time by observing the improvement on the objective function, while the phase shifts of other elements are fixed \cite{Atapattu}. Such element-by-element greedy control is more efficient than convex optimization or ML algorithms with a linear time complexity.

Let us take Fig. \ref{fig-heu}(b) as another example, which applies a greedy policy for RIS element on/off control. Specifically, it applies an element-wise greedy policy and decides the on/off status of one element at each step. Such sequential decision-making leads to a lower time complexity than optimizing all elements simultaneously.  
The last application in Fig. \ref{fig-heu}(c) is to apply the greedy local search for joint active and passive beamforming. Given an initial solution, the local search will find an optimal neighbouring solution that is better than the current solution, and the search is repeated solution-by-solution until it cannot find a better neighbouring solution. 

Despite the low complexity, Fig. \ref{fig-heu}(a) and (c) also reveal that greedy algorithms usually obtain locally optimal results. In Fig. \ref{fig-heu}(a), the greedy decisions (indicated by blue lines) are likely to be different from the globally optimal solution (shown by red lines). Similarly, since the search direction is dominated by the quality of neighbouring solutions, greedy local search in Fig. \ref{fig-heu}(c) may select a wrong searching direction and miss the global optimality.

\subsection{Meta-heuristic Algorithms}
Meta-heuristic algorithms are iterative procedures that utilize specific heuristic strategies to guide the search process. 
As illustrated in Fig. \ref{fig-heu}(d), meta-heuristic algorithms can be classified in various ways, including population-based, trajectory-based, nature-inspired, evolutionary-based, swarm intelligence-based, and so on \cite{ribeiro2007metaheuristics}. For example, genetic algorithm and PSO are well-known evolutionary-based and swarm intelligence-based meta-heuristic algorithms, respectively.

One of the main advantages of meta-heuristic algorithms is the low-design difficulty. Fig. \ref{fig-heu}(e) includes an example of using PSO for RIS phase-shift control. The RIS phase shift is considered as the particle's position, and the optimization objective is defined as the fitness function. In PSO, the movement of each particle is guided by its own best-known position and the entire swarm's best-known position. Then, these particles will constantly update their positions until the best solution converges, achieving a near-optimal RIS phase-shift solution \cite{dai2021reconfigurable}. Similarly, the genetic algorithm applies evolutionary strategies to select elite individual solutions, and produces new solutions by crossover and mutation, guaranteeing that each generation of solutions will be improved until convergence.  

Moreover, the trajectory-based Tabu search is used for joint active and passive beamforming in Fig. \ref{fig-heu}(f). Compared with the greedy search in Fig. \ref{fig-heu}(c), the main difference is that Tabu search can dynamically adjust the search direction by using the Tabu list, which is a set of rules and banned solutions to filter neighbouring solutions. Due to the Tabu rules, Tabu search can explore the solution space more efficiently than greedy search, which is also the main advantage of meta-heuristic algorithms over greedy algorithms. However, meta-heuristic algorithms can be sensitive to key parameter settings, e.g., mutation probability in genetic algorithm and inertia weight in PSO, which have a considerable impact on algorithm performance.

\begin{figure*}[!t]
\centering
\includegraphics[width=0.95\linewidth]{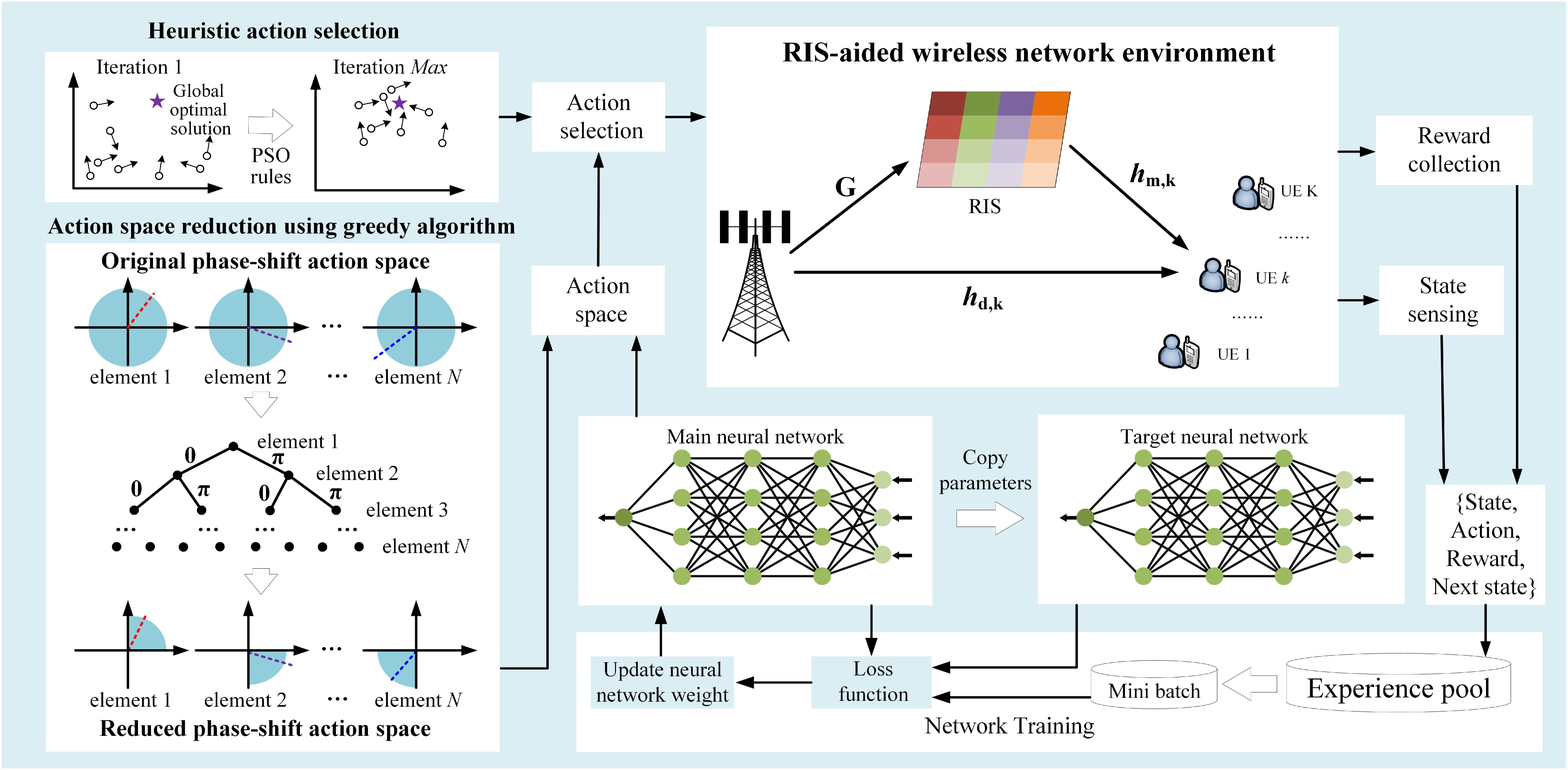}
\caption{Heuristic DRL for RIS phase-shift optimization.}
\label{fig-heudrl}
\vspace{0pt}
\end{figure*}

\subsection{Matching Theory}
Matching theory is particularly useful for resource allocation and association problems, including sub-channel allocation, base station (BS)-RIS-user association, device-to-device (D2D)-user pairing in RIS-aided D2D networks, UAV association in RIS-UAV networks, and so on \cite{bayat2016matching}. These problems are usually NP-hard and cannot be efficiently solved by conventional optimization algorithms. We consider matching theory as a heuristic algorithm because it applies heuristic rules for optimization, which is to find stable matching pairs and then exchange them. 

Matching theory converts the resource allocation into a matching problem, which includes two sets of players with utility functions. 
The core idea of matching theory is to switch matching pairs formed by two sets of players and maximize the total utility function. Consider one set of players are users and another set of players are RISs, and then matching theory is to manipulate the RIS-user associations to maximize the sum-rate or energy efficiency. Specifically, it searches for two exchangeable matching pairs that can improve their utility function by switching the associations, while other players' utility is unaffected. In addition, the association in wireless networks may affect the interference level and network delay. Therefore, the utility function of one player depends not only on its own preference, but also on the matching associations of other users. A similar framework may be applied to RIS-aided D2D and UAV networks as in Fig. \ref{fig-heu}(i), significantly reducing the optimization complexity. However, note that matching theory is an iterative approach to search for matching pairs, and the time cost may exponentially increase when more players are involved.

\section{Heuristic-aided Machine Learning for Optimizing RIS-aided Wireless Networks}

Section \ref{sec-heu} reveals that heuristic algorithms have lower complexity, but the obtained results may be locally optimal. Meanwhile, ML algorithms are the most state-of-the-art techniques for optimizing wireless networks, but tedious model training and slow convergence are well-known issues. Therefore, this section investigates the combination of heuristic algorithms and ML techniques, combining each technique's advantages to obtain low-complexity and high-quality solutions. In the following, we propose three heuristic-aided ML algorithms, namely heuristic DRL, heuristic-aided supervised learning, and heuristic hierarchical learning.
In addition, we analyze heuristic and heuristic-aided ML algorithms in terms of features, advantages, disadvantages, difficulties, and RIS optimization applications.

\begin{figure*}[!t]
\centering
\includegraphics[width=0.95\linewidth]{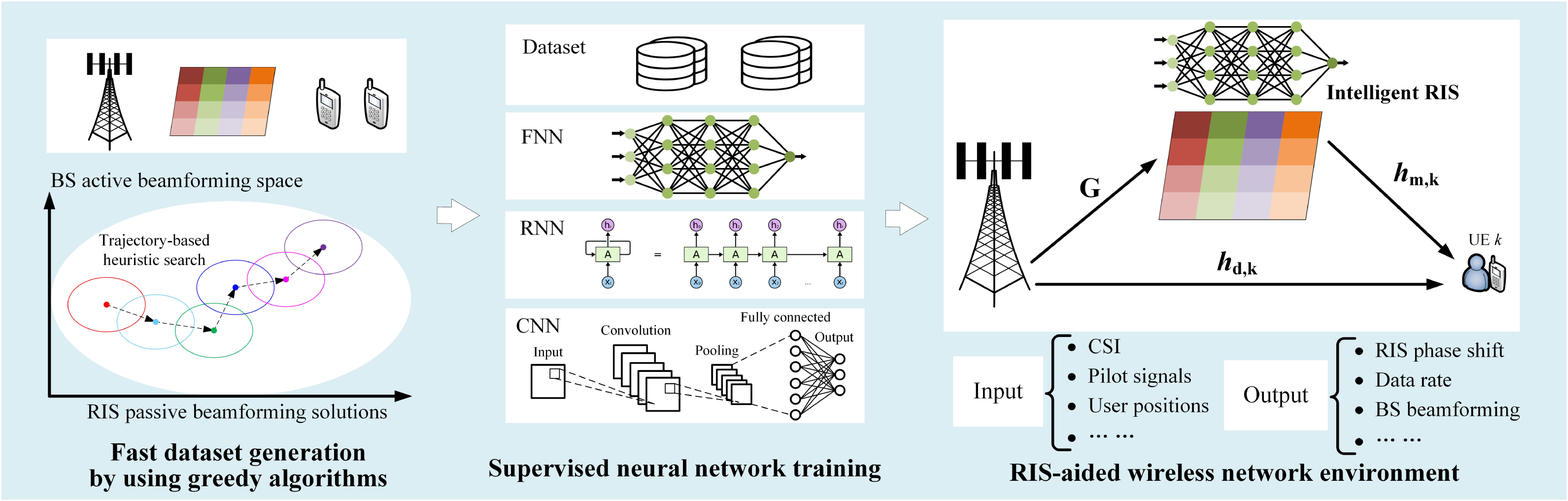}
\caption{Heuristic-aided supervised learning for RIS phase-shift optimization.}
\label{fig-heunn}
\vspace{0pt}
\end{figure*}

\subsection{Heuristic Deep Reinforcement Learning}
DRL is one of the most widely used reinforcement learning techniques for solving optimization problems, and it has been used for optimizing RIS phase shifts in many studies \cite{yliu}. 
However, one of the main difficulties of RIS phase-shift design is the large number of RIS elements, leading to large action spaces for DRL algorithms. 
Subsequently, large action spaces will degrade exploration efficiency and result in sub-optimal results. 
To this end, we integrate heuristic algorithms with DRL to improve exploration efficiency. In particular, as illustrated in Fig. \ref{fig-heudrl}, we first apply greedy algorithms for fast RIS phase-shift selections, which apply an element-by-element greedy policy for phase-shift selection as in Fig. \ref{fig-heu}(a).      
Then, we use the initial results of the greedy algorithm to find the actions that are most frequently visited in each RIS element, which becomes a reduced action space for this element.
The reduced phase-shift space becomes a new action space for DRL, which considerably improves the exploration efficiency.  
Meanwhile, reducing the action space can mitigate the burden of neural network training, since fewer state-action values need to be predicted.
Additionally, heuristic algorithms can also be used for the action selection of DRL. Given current states, meta-heuristic algorithms can be used to search for local optimal actions to maximize the instant reward of the current episode. Compared with random exploration such as $\epsilon$-greedy policy, heuristic exploration can better adapt to a large action space of RIS phase shifts. 

With higher exploration efficiency, heuristic DRL can significantly mitigate model training effort and accelerate convergence. 
Such fast training is critical to capturing RIS-aided network dynamics and thereby making real-time responses. Additionally, by improving the action selection policy, heuristic DRL may achieve higher average rewards, indicating a higher sum-rate or energy efficiency for RIS-aided wireless networks.

However, note that heuristic rules may harm the optimality of DRL, since the potential optimal action may be eliminated by heuristic decisions. In addition, excessive reduction of the action space may result in poor performance. Therefore, the key parameters of heuristic policies in heuristic DRL should be carefully selected, e.g., using the trial and error method.

\subsection{Heuristic-aided Supervised Learning}
Supervised learning is another important approach for optimizing RIS-aided wireless networks. As shown in Fig. \ref{fig-heunn}, it considers various inputs, e.g., CSI, pilot signals, and user positions, and then predicts RIS phase shifts or data rates. 
Supervised learning maps the input to the desired output, but it relies on fine-grained labelled datasets for model training. The accessibility of high-quality datasets may prevent the application of supervised learning, especially for complicated and highly dynamic wireless environments. Many existing studies have used simulators to produce datasets by exhaustive search, but building specific scenarios in the simulator indicates extra complexity\cite{taha2021enabling}. Another approach is applying convex optimization algorithms to produce data\cite{hu2021reconfigurable}. It generates high-quality datasets, but designing convex optimization algorithms requires dedicated analyses.  

Therefore, we propose to apply heuristic algorithms to overcome such dataset availability issues in supervised learning. Specifically, we deploy heuristic algorithms for dataset generation, which will be further used to train the supervised learning models.
Due to the high generalization capability, heuristic algorithms can adapt fast to target scenarios and produce fine-grained datasets. 
Compared with the simulator-based method, the heuristic approach has higher flexibility and dataset quality. Meanwhile, it has lower complexity than using convex optimization to produce datasets. 
Given the dataset, various supervised learning models are selected, such as feedforward neural networks, recurrent neural networks, or convolutional neural networks. Finally, the trained neural networks are used for RIS phase shift or data rate prediction. 
Fig. \ref{fig-heunn} shows that heuristic algorithms can be an appealing approach for generating high-quality datasets, which is crucial for supervised learning applications. Given these datasets, supervised learning models can be better trained, providing more accurate RIS phase-shift or data rate predictions.

\begin{figure*}[!t]
\centering
\includegraphics[width=0.95\linewidth]{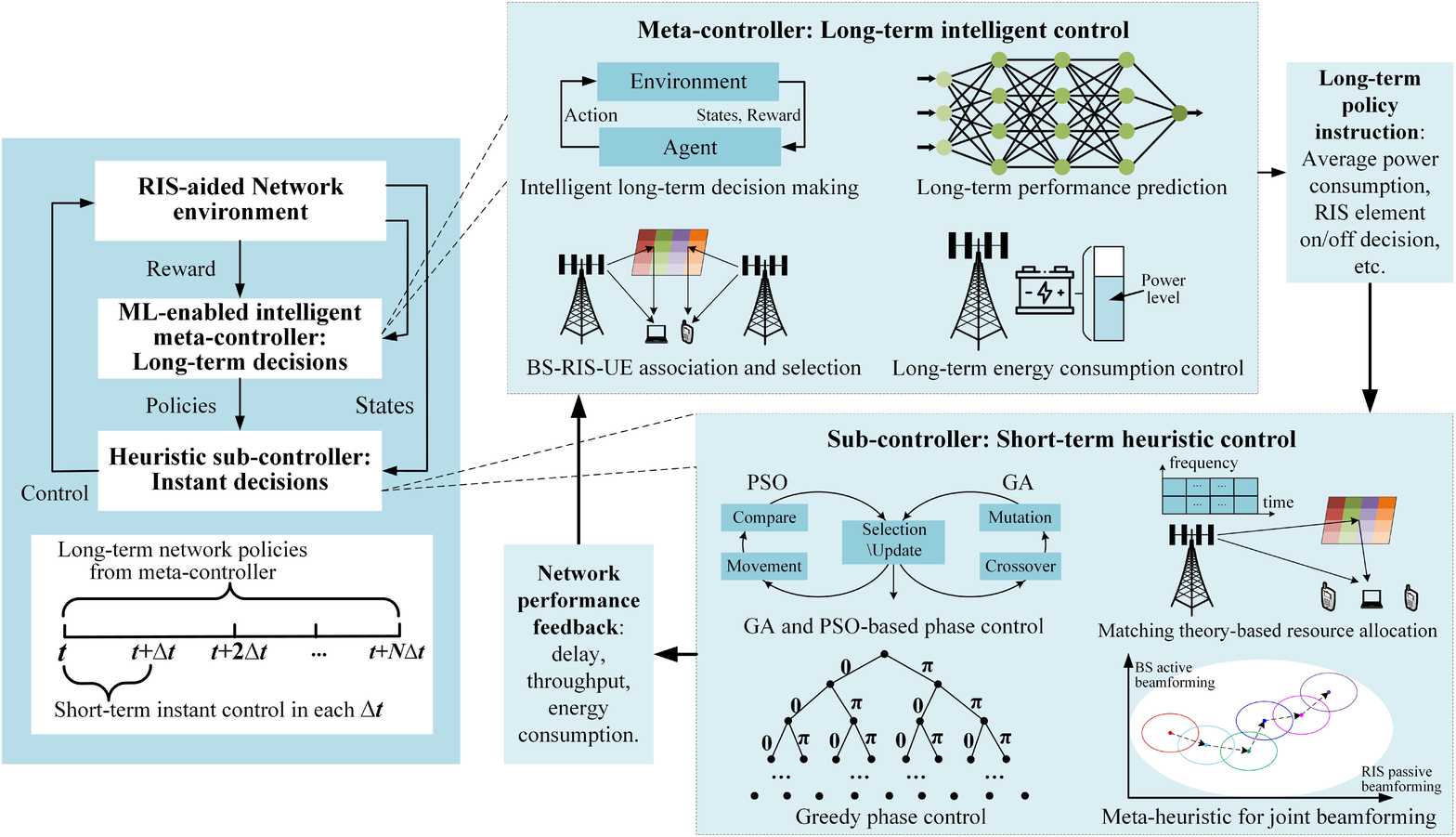}
\caption{Hierarchical framework with heuristic algorithms.}
\label{fig-hrl}
\vspace{0pt}
\end{figure*}

\begin{table*}[!t]
\caption{Summary of heuristic algorithms and heuristic-aided ML for RIS-aided wireless networks }
\centering
\small
\setstretch{1.1}
\resizebox{1\textwidth}{!}{%
\begin{tabular}{|m{1.4cm}<{\centering}|m{3cm}<{\centering}|m{3cm}<{\centering}|m{3.2cm}<{\centering}|m{2.8cm}<{\centering}|m{5.2cm}<{\centering}|}
\hline 
Algorithms  &  Features   &     Advantages    &  Disadvantages   &  Difficulties   &  RIS optimization \qquad \qquad  \qquad \qquad  \qquad \qquad applications  \\
\hline
Greedy algorithms  & Making locally optimal choices at the current stage of solving the problem, and disregarding the following stages.   & Higher flexibility by optimizing the current stage; considerably lower complexity.    &   Greedy algorithms may produce poor results since local optimality cannot guarantee global performance.  &  Balancing optimality and efficiency is difficult when using greedy algorithms.  & \makecell[bl]{1) Greedy element-by-element RIS \\ phase-shift and on/off control;\\ 2) Greedy network scheduling;\\ 3) Greedy local search for joint active \\ and passive beamforming. }      \\
\hline
Meta-heuristic algorithms  &  Applying advanced heuristic policies to search for near-optimal solutions in an iterative manner.   &   Low design complexity; high generalization capability; high-quality solutions.  &  Iterative exploration may be time-consuming; algorithm performance depends on proper parameter selection.  &  Key parameters should be carefully selected, e.g., mutation probability in genetic algorithm.   &  \makecell[cl]{1) Genetic algorithm and PSO for RIS \\ phase-shift optimization;\\  2) Tabu search for RIS element design\\ and on/off control.  } \\
\hline
Matching theory  &   Defining matching pairs to describe two sets of players, maximizing the utility function by switching player associations.  &  Matching theory is easy to implement with a stable output.  &  Matching theory relies on iterative searching, and the complexity may increase when more players are involved.    &  Utility function definition must consider the interference and peer effect in wireless networks.    &  \makecell[bl]{1) Sub-channel assignment;\\ 2) BS-RIS-user association; \\3) UAV-user association in RIS-UAV \\ networks; \\ 4) D2D-user paring in RIS-aided \\ D2D communications.  }  \\
\hline
Heuristic DRL &  Combining heuristic rules with DRL to improve the exploration efficiency.  & Faster convergence and higher average reward than conventional DRL.  & Introducing heuristic rules may harm optimality.  & Incorporating heuristic rules without harming the potential optimality is critical.  & Heuristic DRL is mainly designed for optimization problems with large action space, e.g., RIS phase-shift optimization with discrete phase shifts.  \\
\hline
Heuristic-aided supervised learning & Applying heuristic algorithms to generate datasets for supervised model training.   &  Producing high-quality datasets efficiently for algorithm training.  &  The dataset quality depends on the heuristic algorithm selection, which should be carefully designed.     &  Heuristic algorithm selection for dataset generation.  & Heuristic-aided supervised learning can be used to predict RIS phase shifts, achieved data rates, channel information, and so on.  \\
\hline
Hierarchical learning with heuristic algorithms & Combining hierarchical learning with heuristic algorithms for joint decision-making with various time scales.  &   Enabling hierarchical intelligence for decision-making; bringing long-term benefits for network management.   &  The interaction between the meta-controller and sub-controller may lead to unstable performance.   & Maintaining the stability between the meta-controller and sub-controller can be difficult.   &  Hierarchical management for RIS-aided wireless networks, e.g., the higher level uses DRL for coverage optimization, and the lower level applies greedy algorithms for fast RIS phase-shift control. \\
\hline
\end{tabular}}
\label{tab-overallcom}
\vspace{-10pt}
\end{table*}

\subsection{Heuristic Hierarchical Learning}
RIS phase-shift optimization is usually a short-term instant decision, which requires fast responses to network dynamics. Meanwhile, other network policies may focus on long-term network performance. 
For example, compared with the instant RIS phase-shift optimization, the BS-RIS-user association is usually a long-term decision, or BS sleep control is also a long-term policy that depends on average traffic demand\cite{zhou2023hierarchical}. 
Given such short-term and long-term decisions, coordinating these network functions with different timescales is critical, enabling more flexible and intelligent network management. In addition, such hierarchical intelligence and decision-making may be further generalized to other scenarios such as open radio access network (O-RAN) \cite{zhang2022team}. 

Hence, Fig. \ref{fig-hrl} shows a hierarchical learning framework for the management of RIS-aided networks. The lower layer indicates a sub-controller for instant decision-making in each $\Delta t$, and various heuristic algorithms may be applied, e.g., PSO for RIS beamforming and matching theory for resource allocation. These heuristic algorithms are deployed to produce short-term instant decisions due to their low complexity. Then, the achieved network performance is sent to the ML-enabled meta-controller. Based on the average performance from $t$ to $t+N\Delta t$, the meta-controller applies ML algorithms to decide policies for the next $N\Delta t$, e.g., BS sleep/active status, expected average power consumption level, and required quality of service level. 
ML algorithms can produce more stable results than heuristic algorithms, which fits well with the requirement of long-term network management. Meanwhile, it can adjust the policies dynamically based on the feedback of the sub-controller, and the decision interval $N\Delta t$ provides plenty of time for ML model training.   

Finally, note that RISs are usually jointly optimized with other problems, such as resource allocation and associations, and heuristic hierarchical learning provides a flexible scheme for solving these joint optimization problems. Specifically, short-term and long-term network control decisions can be intelligently and jointly optimized, benefiting the overall performance of RIS-aided wireless networks.

\subsection{Comparison and Analyses}

Table \ref{tab-overallcom} summarizes heuristic algorithms and heuristic-aided ML techniques, including features, advantages, disadvantages, difficulties and RIS optimization applications. 

Greedy algorithms make locally optimal choices at the current step of solving the problem, disregarding the effect on the following steps. Therefore, it has lower complexity but may produce poor results, since local optimality in each step cannot guarantee global optimality. Greedy algorithms are widely used for RIS phase-shift design and network management as low-complexity alternatives, e.g., element-by-element RIS phase-shift optimization and on/off control. 

Compared with greedy algorithms, meta-heuristic algorithms include more advanced policies to search for near-optimal solutions. The main advantage is the low design complexity, which indicates that control variables, constraints, and objectives can be easily transformed into the meta-heuristic context. Heuristic rules, such as evolutionary policies in the genetic algorithm, can guarantee the quality of solutions. However, some key parameters, e.g., mutation probability in the genetic algorithm, may greatly affect the algorithm performance, which should be carefully selected.  

Matching theory specializes in solving resource allocation and association problems.
These NP-hard allocation problems are transformed into matching pairs, and then improve the utility function by switching pairs. Matching theory is easy to implement with stable output, but it relies on extensive searches to improve the matching associations between players. Hence, matching theory may be time-consuming when involving many players, e.g., BS-RIS-user association problems with many users.     

For heuristic-aided ML algorithms, heuristic DRL is expected to achieve faster convergence than conventional DRL by reducing the action space and heuristic exploration. 
However, one potential disadvantage is that the optimality may be degraded, since heuristic DRL trades optimality for efficiency and convergence. Subsequently, how to apply heuristic rules to DRL without undermining optimality can be challenging. 
Compared with conventional supervised learning, the main difference of the proposed heuristic supervised learning is to utilize heuristic approaches to generate high-quality datasets. It indicates that fine-grained datasets can be easily produced for specific scenarios, lowering the difficulty of supervised learning deployment. 
Meanwhile, hierarchical learning provides a framework for joint network management. It coordinates short-term and long-term network control, enabling a more flexible management scheme. However, the stability between the meta-controller and sub-controller should be carefully maintained.    

Finally, the former sections focus on RIS phase-shift design because this is the core control variable for optimizing RIS-aided networks. Additionally, other variables may also be involved, e.g., BS transmit power control and resource allocation, resulting in more complicated joint optimization problems. To this end, one may apply greedy policies to optimize each variable sequentially, lowering the difficulty of optimizing all sub-problems simultaneously. 
For heuristic-aided supervised learning, more advanced neural network models can be deployed to handle the increasing number of input variables.
For joint optimization problems with hierarchies, hierarchical learning can also be applied to decouple the joint optimization problem into several sub-problems.

\begin{figure*}[!t]
\centering
\subfigure[Sum-rate performance of various algorithms.]{ \label{fig_rate}
\includegraphics[width=8cm,height=6cm]{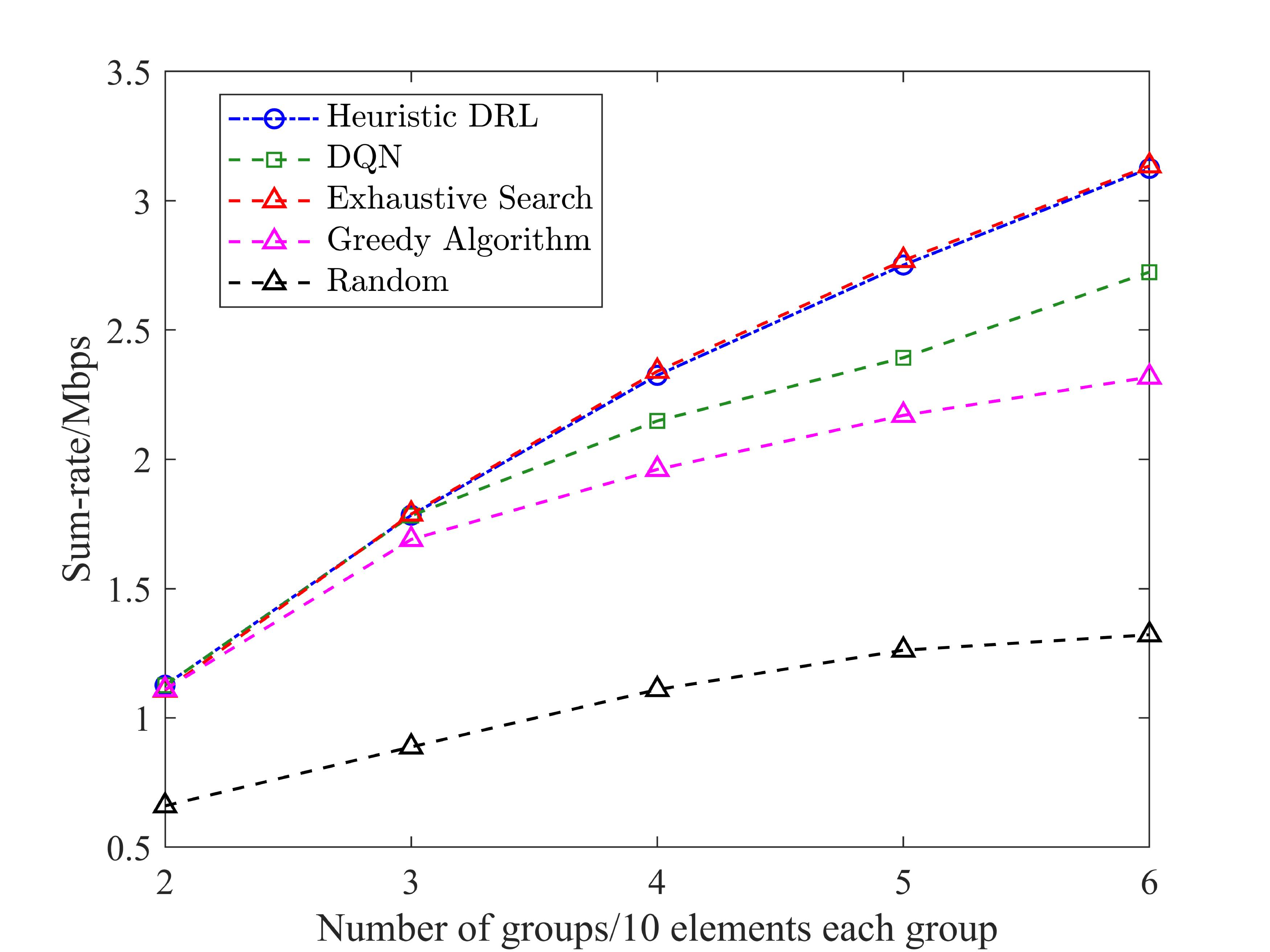}
}
\,
\subfigure[Convergence comparisons of DQN and heuristic DRL.]{ \label{fig_heu_com}
\includegraphics[width=8cm,height=6cm]{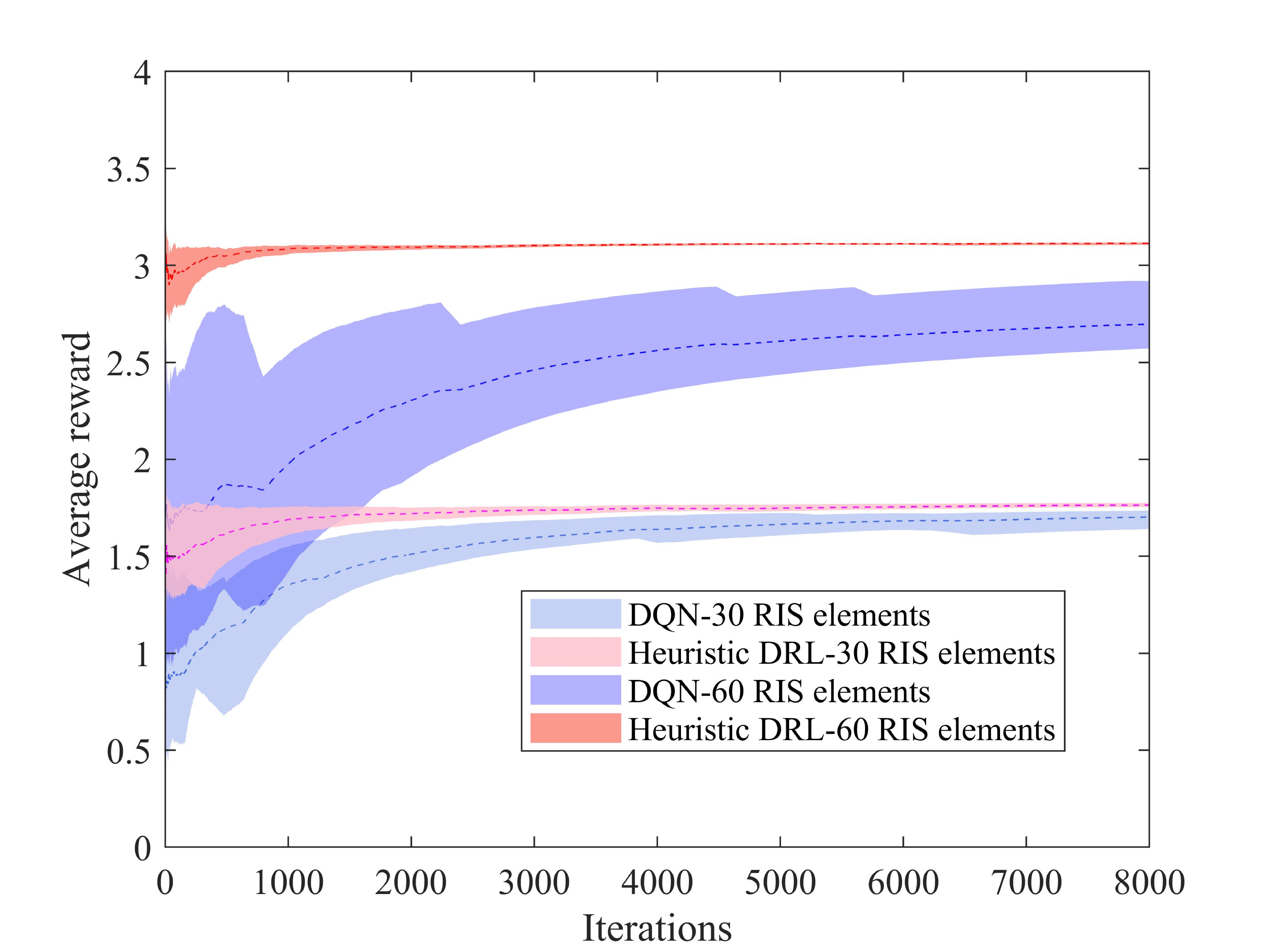}
}
\subfigure[Runtime complexity comparison (Simulation time of 1 run with 8000 iterations).]{ \label{fig_simtime}
\includegraphics[width=7.3cm,height=5.8cm]{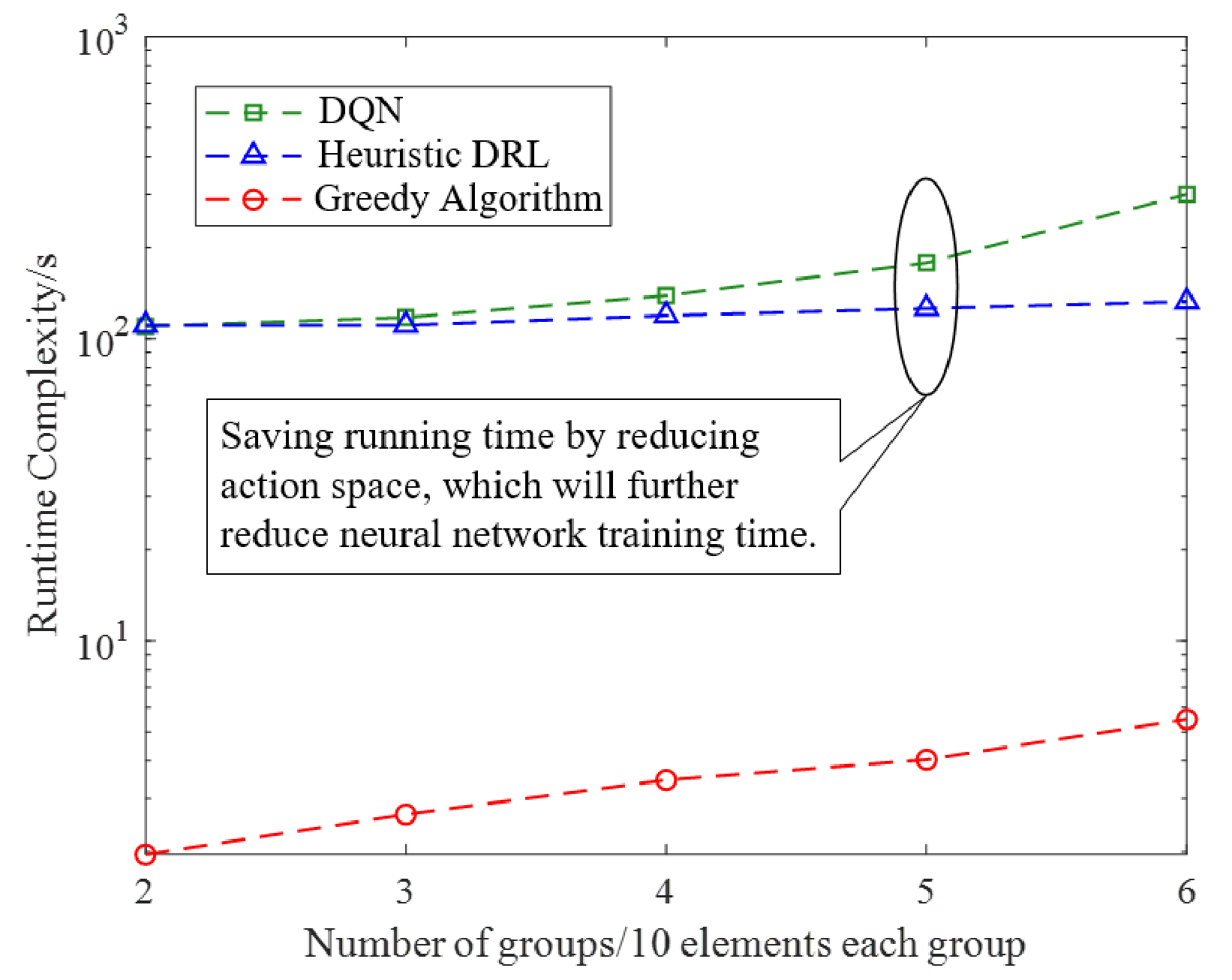}
}
\,
\subfigure[Sum-rate under different percentage of action space reduction in heuristic DRL.]{  \label{fig_trade}
\includegraphics[width=8cm,height=6cm]{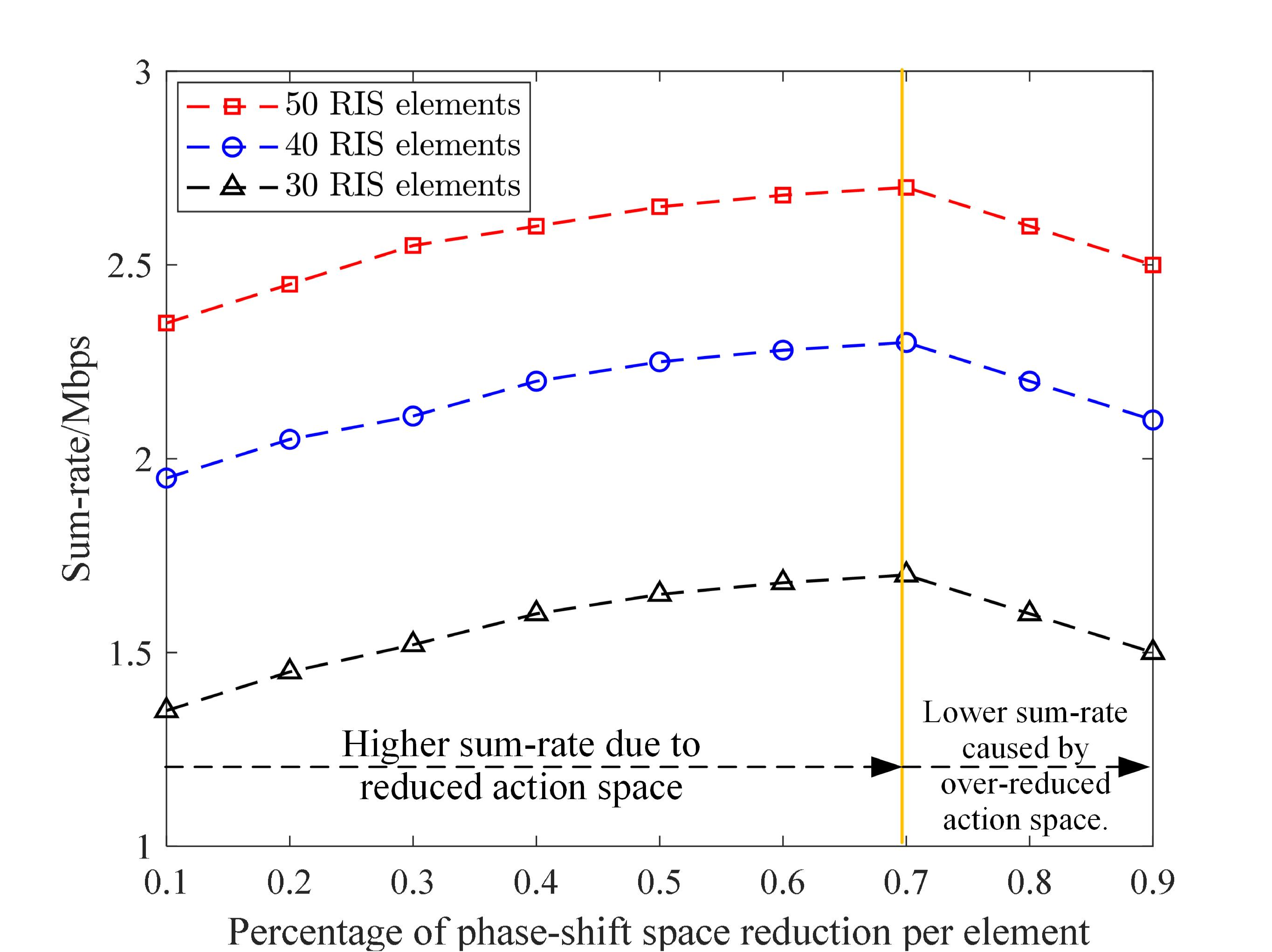}
}
\setlength{\abovecaptionskip}{0pt} 
\caption{Simulation results comparison }
\vspace{-15pt}
\label{fig-resu}
\end{figure*}

\section{ Case Study on RIS Phase-shift Control}
This section presents a case study on RIS phase-shift optimization with one multi-antenna BS and 5 single-antenna users. We assume that the direct link between BS and users is blocked by obstacles, and the BS-RIS-user channel follows the Rician fading. The BS-RIS distance is 50 m, and the RIS-user distance is randomly distributed between 50 to 60 m. We describe channels in the Rician model, which is widely used to characterize the channel models in RIS-aided wireless networks. The BS-RIS-user links act as the dominant line-of-sight component and all the other paths contribute to non-line-of-sight components.
We consider 10 RIS elements with 2 bits as a group to ease the computational complexity, which means each group makes the same phase-shift decision\cite{cao2021reconfigurable}. Here 2-bit resolution means 4 possible phase changes: 0, $\frac{\pi}{2}$, $\pi$, $\frac{3\pi}{2}$. Grouping means that adjacent RIS elements with high channel correlation are considered as a sub-surface, which will share the same reflection coefficient. The complexity of reflection design can be significantly reduced by grouping, which is a widely considered approach. The simulations involve 4 algorithms, namely heuristic DRL, conventional deep Q-networks (DQN), exhaustive search, and random phase shift. 
Heuristic DRL has been introduced in Fig. \ref{fig-heudrl}, which applies greedy element-by-element phase-shift optimization to reduce the action space. Exhaustive search is the optimal baseline, while random phase shift means selecting phase shifts randomly. We select DQN as a baseline since it is the most widely applied reinforcement learning algorithm for optimization, and it has been used in a few studies for RIS phase-shift design\cite{kfai}.
The simulation is implemented in MATLAB for 10 runs, and the average results are shown in Fig. \ref{fig-resu}.

Fig. \ref{fig_rate} shows the sum-rate of all users under various algorithms and RIS numbers. When the number of RIS elements is 20 or 30, the action space is still small, and both DQN and heuristic DRL can explore different phase-shift combinations and then find the optimal action. Therefore, DQN and heuristic DRL achieve comparable performance as the exhaustive search method, and the random strategy shows the worst performance. However, when the number of RIS elements increases to 40, the action space will exponentially expand. The DQN agent has to explore hundreds of actions, and it is difficult to attain good policies with such a limited number of iterations, leading to degraded performance.  
In contrast, heuristic DRL applies greedy algorithms to reduce the action space, and then the compressed action space is explored efficiently. As a result, heuristic DRL still maintains a comparable sum-rate as the exhaustive search method.

Moreover, Fig. \ref{fig_heu_com} compares the convergence of DQN and heuristic DRL, demonstrating that heuristic DRL obtains higher average reward and faster convergence than DQN. The main reason is that the actions of heuristic DRL have been preselected by heuristic algorithms, indicating higher exploration efficiency. 
On the contrary, the DQN agent has to explore hundreds of actions randomly, leading to a slower convergence and lower average rewards. 
Fig. \ref{fig_simtime} presents the runtime complexity of heuristic DRL, DQN, and the greedy algorithm. Note that the simulation time may be affected by the hardware settings of simulation platforms, and our simulation platform is MATLAB with Intel core i7-7770 CPU with 16 G memory size. Fig. \ref{fig_simtime} reveals a clear trend that heuristic DRL can save up to 60\% lower simulation time than DQN. One of the most time-consuming parts of running DRL is the neural network training. Reducing action spaces can considerably lower the burden of neural network training, since fewer state-action pairs need to be learned and predicted. In addition, the greedy algorithm has a much lower runtime complexity than heuristic DRL and DQN, which is one of the main advantages of heuristic algorithms. Such low complexity enables rapid response to RIS-aided network dynamics.

Finally, Fig.\ref{fig_trade} shows the performance trade-off of the proposed heuristic DRL algorithm. Specifically, it demonstrates that the percentage of phase-shift space reduction may affect the algorithm performance. An oversimplified action space will undermine the potential optimality and lead to a lower sum-rate, while the system performance improvements are less obvious with a slight action space reduction.
The optimal space reduction percentage is around 0.7, which is used for obtaining simulation results in \ref{fig_rate} and \ref{fig_heu_com}. 
In summary, the simulation results in Fig. \ref{fig-resu} demonstrate that integrating heuristic algorithms with ML algorithms can achieve higher data rates and faster convergence. These combinations can make full use of each technique's advantage, providing a new perspective for optimizing RIS-aided wireless networks.

\section{Future Directions of Heuristic Algorithms for RIS-aided Wireless Networks}

\textbf{1) Investigating Performance Limits of Heuristic ML}:
There are several benefits of incorporating heuristic algorithms into ML. However, the inherent features of heuristic policies may undermine the stability of ML algorithms. Therefore, it is crucial to identify the performance limits of heuristic-aided ML algorithms. Such an investigation will facilitate the application of heuristic-aided ML techniques.

\textbf{2) Greedy Policy for ML}:
Greedy policies can be very useful in improving the efficiency of ML techniques. For example, the epsilon-greedy policy has been widely used to balance the exploration and exploitation of reinforcement learning algorithms. In this work, we propose to use the greedy policy to reduce the action space of RIS phase-shift designs. It is expected that introducing greedy policies into ML can bring more efficient techniques for the control and optimization of RIS-aided wireless networks.

\textbf{3) Population-based Heuristic ML}:
Population-based heuristic algorithms, such as the genetic algorithm and PSO, have been successfully applied to many problems, enabling population intelligence to attain near-optimal solutions. Such population-based intelligence can be an appealing approach to solve large-scale optimization problems. For instance, letting each intelligent agent be an individual in the genetic algorithm, and deploying evolutionary strategies to update and guide individuals. However, this requires research efforts in terms of algorithm design and implementation.

\textbf{4) Heuristic-aided Lightweight ML}:
Wireless networks are becoming increasingly complicated due to new architectures and designs, requiring larger scale ML algorithms, e.g., more neural network nodes and layers. However, this also leads to considerable algorithm training and tuning costs. Heuristic algorithms provide opportunities to develop lightweight ML techniques. Specifically, this enables the use of small-scale ML techniques with fewer algorithm parameters and lower training costs, which fits well with resource-constrained devices in RIS-aided wireless networks.

\textbf{5) Heuristic-aided ML for Joint Optimization Problems}
Joint optimization problems are very common in wireless networks, involving multiple highly-coupled control variables. One conventional approach is to jointly transform all control variables into an action, and then use reinforcement learning for optimization. However, this will lower the exploration efficiency and result in sub-optimal results. To this end, one may decouple the joint optimization into multiple stages, and optimize each stage in a greedy manner for lower complexity.

\section{Conclusion}
It is anticipated that RIS technology may be a key enabler for 6G networks, and our work has investigated heuristic algorithms and their applications for RIS-aided wireless networks, including greedy algorithms, meta-heuristic algorithms, and matching theory. In addition, we have integrated heuristic algorithms with ML techniques, namely heuristic-aided ML. The case study shows that heuristic-aided ML can significantly improve the convergence of DRL and achieve higher data rates. 
In the future, we plan to study the performance of heuristic-aided hierarchical learning, and further integrate convex optimization, heuristic algorithms, and ML techniques for joint network management. 

\section*{Acknowledgment}

This work was supported in part by the Canada Research Chairs Program and the U.S National Science Foundation under Grant CNS-2128448 and ECCS-2335876.

\normalem

\begin{IEEEbiography}
{Hao Zhou} obtained his PhD degree at the University of Ottawa in 2023. Before this, he got his M.Eng degree from Tianjin University, China, in 2019. His research interests include microgrid energy trading, O-RAN, resource management and network slicing. He is devoted to developing machine learning techniques for smart grid and 5G/6G applications. He received the best paper award at the 2023 IEEE ICC conference, and won the Outstanding Self-financed Abroad Chinese Students Award.    
\end{IEEEbiography}

\begin{IEEEbiography}
{Melike Erol-Kantarci} is Chief Cloud RAN AI ML Data Scientist at Ericsson and Canada Research Chair in AI-enabled Next-Generation Wireless Networks and Full Professor at the University of Ottawa. She is the co-editor of three books on smart grids and intelligent transportation. She has over 200+ peer-reviewed publications with citations over 8000 and h-index 43. Throughout her career, she has received numerous awards and recognitions; has delivered numerous keynotes, plenary talks and tutorials around the globe both at industry events and academic conferences. Her main research interests are AI-enabled wireless networks, 5G and 6G wireless communications and smart grids. She is an IEEE ComSoc Distinguished Lecturer and ACM Senior Member.
\end{IEEEbiography}

\begin{IEEEbiography}
{Yuanwei Liu} (Senior Member, IEEE, \text{http://www.eecs.qmul.ac.uk/} {$\sim$ yuanwei}) is a senior lecturer at Queen Mary University of London, London, E1 4NS, U.K. His research interests include NOMA, RIS/STAR, Integrated Sensing and Communications, Near-Field Communications, and machine learning. He is a Web of Science Highly Cited Researcher since 2021, an IEEE Communication Society Distinguished Lecturer, an IEEE Vehicular Technology Society Distinguished Lecturer, and the academic Chair for the Next Generation Multiple Access Emerging Technology Initiative. He also serves as an Editor-in-Chief of IEEE ComSoc TC Newsletter, an Area Editor of IEEE CL, and an Editor of the IEEE COMST/TWC/TVT/TNSE.
\end{IEEEbiography}

\begin{IEEEbiography}
{H. Vincent Poor} is the Michael Henry Strater University Professor at Princeton University, where his interests include information theory, machine learning and network science, and their applications in wireless networks, energy systems, and related fields. He is a member of the U.S. National Academy of Engineering and the U.S. National Academy of Sciences, and a foreign member of the Royal Society and other national and international academies. He received the IEEE Alexander Graham Bell Medal in 2017.
\end{IEEEbiography}

\end{document}